\documentclass{PoS}

\usepackage{float}
\usepackage{url}

\newtheorem{theorem}{Theorem}

\title{Mapping theorem and Green functions in Yang-Mills theory}

\ShortTitle{Mapping theorem and Yang-Mills theory}

\author{\speaker{Marco Frasca} \\
        Via Erasmo Gattamelata, 3\\
        00176 Roma (Italy) \\
        E-mail: \email{marcofrasca@mclink.it}}

\abstract{It is shown how, starting from a mapping theorem recently proved between massless quartic scalar field theory and Yang-Mills theory, both two-point functions and spectrum of the Yang-Mills theory can be obtained. These results compare very well with respect to lattice computations.}

\FullConference{The many faces of QCD\\
		November 1-5, 2010\\
		Gent Belgium}

\begin{document}

\section{Introduction}

Since Lattice 2007 conference in Regensburg, there has been a paradigm shift in the study of propagators of the Yang-Mills theory in the Landau gauge \cite{ilg,cuc,oli}. Before that date it was generally accepted that the gluon propagator in Landau gauge should have gone to zero at lower momenta with the ghost propagator going to infinity faster than the free case and the running coupling reaching a non-trivial fixed point at infrared \cite{alk}. Lattice computations performed at very large volumes and reaching the considerable dimension of $(27fm)^4$ \cite{cuc} proved on the contrary that the scenario to be considered is one having the gluon propagator reaching a finite non-zero value with a ghost propagator behaving as that of a free particle and the running coupling, once a definition in the infarerd is agreed, reaching a trivial infrared fixed point. This latter solution was named the ``decoupling solution'' to distinguish it from the one firstly assumed true that then was called ``scaling solution''. 

The decoupling solution was postulated as early as in the eighties by Cornwall \cite{corn} and obtained in the course of time by several other authors \cite{an,ap,ab,bou,fra1,fra2}. A preferred technique in this kind of studies has been solving Dyson-Schwinger equations numerically but, quite recently, we have been able to provide a strong coupling expansion in quantum field theory \cite{fra1} and then a theorem mapping the behavior of Yang-Mills theory on that of a massless scalar field theory at infrared \cite{fra2,fra3}. This technique opened up the opportunity to get explicit perturbative solutions in the formal limit of the coupling going to infinity exploiting completely the behavior of the Yang-Mills theory at low energies. This contribution gives an account of this approach and how it nicely fits into the scenario so far emerged both from lattice computations and other theoretical analysis.

\section{Classical field theories}

It is a common way to give an understanding of quantum field theory to start from solutions of the corresponding classical theory. This provides a set of solutions to start from. Here we will follow this track and we start by considering the equation of motion of a massless scalar field
\begin{equation}
\label{eq:phi}
   \Box\phi+\lambda\phi^3=j.
\end{equation}
Our aim is to give an expression for the solution functional $\phi[j]$ that holds in the limit $\lambda\rightarrow\infty$. But we can give immediately a solution to the corresponding homogeneous equation. We have
\begin{equation}
   \phi(x)=\mu\left(\frac{2}{\lambda}\right)^\frac{1}{4}{\rm sn}(p\cdot x+\theta,i)
\end{equation}
being sn a Jacobi elliptic function with elliptic modulus $i$, $\theta$ and $\mu$ two arbitrary integration constants. This solution holds provided
\begin{equation}
   p^2=\mu^2\left(\frac{\lambda}{2}\right)^\frac{1}{2}
\end{equation}
that is the dispersion relation of a massive free particle. So, all the effect of the nonlinear term in the equation is to produce mass to the field. This is a dynamical effect and can be seen as the classical analogous of the Schwinger mechanism. It is important to emphasize that if we consider just the limit of a small coupling $\lambda$ and do perturbation theory we are not able to recover this solution, rather we could have to cope with some singularities in our expansion. This means that this solution could be only unveiled if we would be able to do a perturbation expansion in a strong coupling limit \cite{fra4}. In order to reach our aim for eq.(\ref{eq:phi}) we look for a solution in the form
\begin{equation}
   \phi(x)=\mu\int d^4x'G(x-x')j(x')+\delta\phi(x).
\end{equation} 
being $\mu$ the same constant obtained above. One can see that is is so provided \cite{fra5}
\begin{equation}
\label{eq:green}
   \delta\phi(x)=\mu^2\lambda\int d^4x'd^4x''G(x-x')[G(x'-x'')]^3j(x')+O(j^3)
\end{equation}
and $\Box G(x-x')+\lambda[G(x-x')]^3=\mu^{-1}\delta^4(x-x')$ the Green function of the theory for finite $\lambda$. So, in order to completely solve the classical equation of motion we need to know $G$. This can be done in a straightforward way by consider the theory in d=1+0 and solving for $\partial_t^2G_0(t-t')+\lambda [G_0(t-t')]^3=\mu^2\delta(t-t')$ \cite{fra1} and so
\begin{equation}
   G_0(\omega)=\sum_{n=0}^\infty(2n+1)\frac{\pi^2}{K^2(i)}\frac{(-1)^{n}e^{-(n+\frac{1}{2})\pi}}{1+e^{-(2n+1)\pi}}
   \frac{1}{\omega^2-m_n^2+i\epsilon}
\end{equation}
being $m_n=(2n+1)\frac{\pi}{2K(i)}\left(\frac{\lambda}{2}\right)^{\frac{1}{4}}\mu$ and $K(i)\approx 1.3111028777$ an elliptic integral. Then, after a Lorentz boost we get finally the full Green function
\begin{equation}
   G(p)=\sum_{n=0}^\infty(2n+1)\frac{\pi^2}{K^2(i)}\frac{(-1)^{n}e^{-(n+\frac{1}{2})\pi}}{1+e^{-(2n+1)\pi}}
   \frac{1}{p^2-m_n^2+i\epsilon}.
\end{equation}
From the way it depends on $\lambda$ we see that we have a strong coupling expansion for the solution of the theory that holds for $\lambda\rightarrow\infty$.

We can work with an identical technique also for a classical Yang-Mills theory. Given the equations of motion
\begin{equation}
\partial^\mu\partial_\mu A^a_\nu-\left(1-\frac{1}{\alpha}\right)\partial_\nu(\partial^\mu A^a_\mu)+gf^{abc}A^{b\mu}(\partial_\mu A^c_\nu-\partial_\nu A^c_\mu)+gf^{abc}\partial^\mu(A^b_\mu A^c_\nu)+g^2f^{abc}f^{cde}A^{b\mu}A^d_\mu A^e_\nu = -j^a_\nu
\end{equation}
being as usual $g$ the coupling and $\alpha$ the contribution of the gauge fixing term. Now we note that, for the homogeneous equation, a set of exact solutions can always be found, all depending just on the time variable, if we select some components of the potential and we take all of them equal \cite{smi}. These solutions are just replicas of the exact solution of the scalar field in the rest frame. But we cannot just boost these solutions to get exact ones as Yang-Mills theory has another symmetry playing a role, the gauge symmetry, and we can only hope in a set of approximate solutions that at leading order can map the scalar field solution. In the general case we can always check the same technique applied to the scalar field and write down
\begin{equation}
\label{eq:ympot}
   A_\mu^a=\Lambda\int d^4x'D_{\mu\nu}^{ab}(x-x')j^{b\nu}(x')+\delta A_\mu^a.
\end{equation}
Indeed, also in this case is possible an expansion in powers of the currents as already guessed in the eighties \cite{rob}. So, we are arrived to an identical stumbling block as the one met in the eighties: The correct expression for the gluon propagator in the infrared limit \cite{gol,rob}. The propagator of the gluon field can be immediately obtained through the following mapping theorem that shows how, in the low energy limit, both the massless scalar field and the Yang-Mills theory map producing massive solutions already at classical level. We have proved the following {\sl mapping theorem} \cite{fra2,fra3}

\begin{theorem}
An extremum of the action
\begin{equation}
\nonumber
    S = \int d^4x\left[\frac{1}{2}(\partial\phi)^2-\frac{\lambda}{4}\phi^4\right]
\end{equation}
is also an extremum of the SU(N) Yang-Mills Lagrangian when one properly chooses $A_\mu^a$ with some components being zero and all others being equal, and $\lambda=Ng^2$, being $g$ the coupling constant of the Yang-Mills field, when only time dependence is retained. In the most general case the following mapping holds
\begin{equation}
\nonumber
    A_\mu^a(x)=\eta_\mu^a\phi(x)+O(1/\sqrt{N}g),
\end{equation}
being $\eta_\mu^a$ some constants properly chosen, that becomes exact for the Lorenz gauge.
\end{theorem}

\begin{figure}[H]
\begin{center}
\includegraphics[width=400pt,height=120pt]{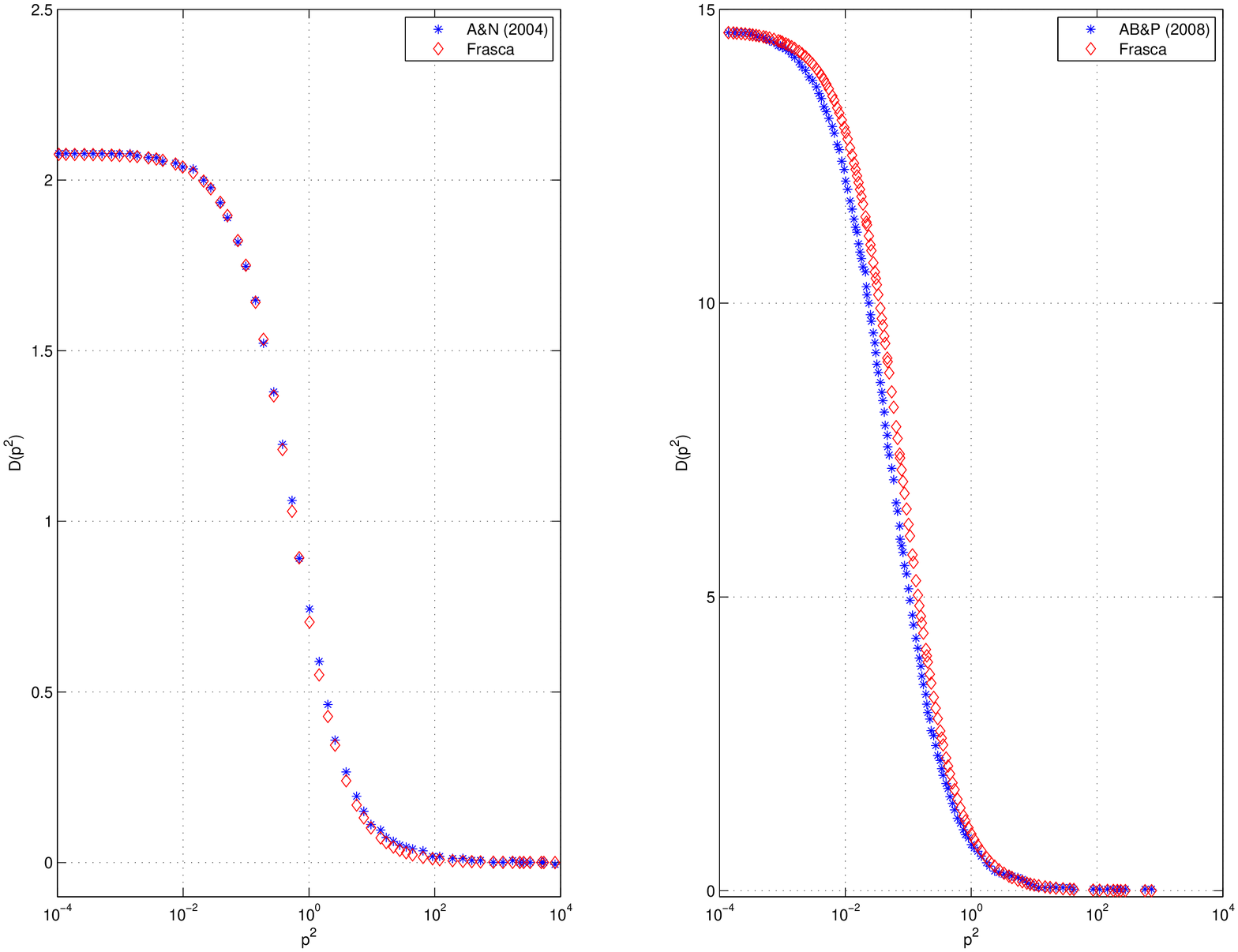}
\caption{\label{fig:fig1} Numerical solutions of Dyson-Schwinger equations: (left) Ref.\cite{an}; (right) Ref.\cite{abp}.}
\end{center}
\end{figure}

This theorem was definitively proved in Ref.\cite{fra3} after a criticism by Terence Tao. Tao agreed with the correctness of this latter proof \cite{tao}. So, we can write down immediately the gluon propagator in the Landau gauge. One has
\begin{equation} 
\label{eq:ymprop}   
  \Delta_{\mu\nu}^{ab}(p)=\delta_{ab}\left(\eta_{\mu\nu}
  -\frac{p_\mu  p_\nu}{p^2}\right)\sum_{n=0}^\infty\frac{B_n}{p^2-m_n^2+i\epsilon}
  +O\left(\frac{1}{\sqrt{N}g}\right)
\end{equation}
being
\begin{equation}
    B_n=(2n+1)\frac{\pi^2}{K^2(i)}\frac{(-1)^{n+1}e^{-(n+\frac{1}{2})\pi}}{1+e^{-(2n+1)\pi}}
\end{equation}
and
\begin{equation}
    m_n=(2n+1)\frac{\pi}{2K(i)}\left(\frac{Ng^2}{2}\right)^{\frac{1}{4}}\Lambda.
\end{equation}
It should be emphasized that the constant $\Lambda$ is the same seen in the high-energy limit for QCD. Here appears as an integration constant \cite{fra6}. A couple of comments are in order. We see from this propagator that the leading order in a strong coupling expansion for a classical Yang-Mills theory is made by a free massive field: This is a classical version of the Schwinger mechanism at work. Mass arises from a finite nonlinearity in the equations of the field. So, it is surely a nontrivial task to solve also the classical equations on a lattice provided we consider the interesting limit of a finite coupling. The next question to be asked is if this is the right propagator to use for the corresponding quantum Yang-Mills theory. To answer this question we compared this propagator with the results obtained solving numerical Dyson-Schwinger equations \cite{an,abp}. The result is given in fig.\ref{fig:fig1}.
The only fitting parameter has been the gluon mass fixed to 737 Mev and 278 MeV respectively. The agreement is really striking. We just note a small discrepancy in the intermediate regime but this should be expected as our propagator is just an approximate one that holds in the deep infrared limit. Indeed, as we will see below, this is the propagator of a free theory with a massive particle and its excited states. Higher order corrections are expected to modify this propagator producing a momenta dependent mass improving the agreement also in the intermediate regime as also put forward by Cornwall in the eighties \cite{corn}. Indeed, a recent lattice analysis \cite{oli2} shows that a good fitting with a propagator like the one given here should be expected to require at least four masses, very similar to the physical masses of glueballs obtained on the lattice \cite{tep,morn}. This agreement improves taking a mass depending on momenta as $m^2+cp^2$ \cite{oli2}.

A mass depending on momenta should be expected by accounting for higher order corrections \cite{fra5} in the same form as given in \cite{oli2}. The theory is renormalized fixing the cut-off through the physical constant $\Lambda$. Higher order corrections should improve the situation giving a gluon propagator fitting on all the range of energies in agreement with the one given by Cornwall \cite{corn}.

\section{Quantum field theory}

In order to formulate a strong coupling expansion for quantum field theory of the scalar field we start with the following generating functional
\begin{equation}
    Z[j]=N\int[d\phi]\exp\left[i\int d^4x\left(\frac{1}{2}(\partial\phi)^2-\frac{\lambda}{4}\phi^4+j\phi\right)\right].
\end{equation}
We want to evaluate it in the formal limit $\lambda\rightarrow\infty$. In order to do this and to exploit our ability to solve the classical theory in the same limit we do the rescaling of variables $x\rightarrow\sqrt{\lambda}x$.
\begin{equation}
    Z[j]=N\int[d\phi]\exp\left[i\frac{1}{\lambda}\int d^4x\left(\frac{1}{2}(\partial\phi)^2-\frac{1}{4}\phi^4+\frac{1}{\lambda}j\phi\right)\right].
\end{equation}
Now, if we take the solution series $\phi=\sum_{n=0}^\infty\lambda^{-n}\phi_n$ and rescale the current $j\rightarrow j/\lambda$ being this arbitrary, it is easy to see that the leading order is obtained by solving the equation of the classical theory $\Box\phi+\lambda\phi^3=j$. But we already know how to manage this equation, as discussed in the preceding section, and a strong coupling expansion is promptly given. We point out that
this is completely consistent with our preceding formulation \cite{fra1} but now the treatment is fully covariant. We are just using our ability to solve the classical theory. Now, the proof of triviality in the infrared of the scalar field theory is just a small step away. We recall eq.(\ref{eq:green}) and we get at the leading order the following Gaussian generating functional
\begin{equation}
    Z_0[j]=Z[0]\exp\left[\frac{i}{2}\int d^4x'd^4x''j(x')G(x'-x'')j(x'')\right].
\end{equation}
This result is really important because gives an explicit demonstration that, in the infrared limit, the four-dimensional massless scalar field theory reaches a trivial fixed point but the spectrum is that of a massive particle, i.e. the theory displays a mass gap, with all its excitations with the spectrum of a harmonic oscillator
\begin{equation}
    m_n=(2n+1)\frac{\pi}{2K(i)}\left(\frac{\lambda}{2}\right)^\frac{1}{4}\mu.
\end{equation}
These are just the poles of the propagator of the classical theory and we will see below why they can be interpreted in this way. If we force this result pretending that K\"allen-Lehman representation holds, this says to us that, at the trivial infrared fixed point, the theory has no bound states but just free massive particles and their excited states. These states in turn can be used as asymptotic states to build the low-energy limit of the theory.

We can take all this to the case of a quantum Yang-Mills theory. So, by using the propagator (\ref{eq:ymprop}), or in some other gauge by using the mapping theorem, and the change of variables (\ref{eq:ympot}), we get for the generating functional
\begin{equation}
   Z[j]=Z[0]\exp\left[\frac{i}{2}\int d^4x'd^4x''j^{a\mu}(x')D_{\mu\nu}^{ab}(x'-x'')j^{b\nu}(x'')+O(j^3)\right]+O\left(\frac{1}{\sqrt{N}g}\right).
\end{equation}
This functional has a Gaussian form and describes a theory of free massive particles and their excitation spectrum. This implies that the theory reaches an infrared trivial fixed point. These particles are the asymptotic states to build upon a quantum theory in the infrared limit for the Yang-Mills field. As such, these should be observable states that can be termed "glueballs''. This holds in the limit of the coupling going to infinity describing the physics of Yang-Mills field at the infrared fixed point. If we look at this generating functional we note that the ghost field disappeared. For the given propagator, with our change of variables in the path integral, the ghost field just decouples and we can write down its propagator as
\begin{equation}
   G_{ab}(p)=-\frac{\delta_{ab}}{p^2+i\epsilon}+O\left(\frac{1}{\sqrt{N}g}\right).
\end{equation}
The next to leading term was computed on the lattice by Cucchieri and Mendes at $\beta=0$ \cite{cuc2}. From this we can conclude that, in a strong coupling expansion in the inverse of the 't Hooft coupling, we get the decoupling solution.

We can draw some conclusions also about the running coupling. We note that we have derived the generating functional at the infrared fixed point where it takes the Gaussian form, i.e. the theory is trivial there. Using the standard definition of an infrared running coupling \cite{ste}
\begin{equation}
     \alpha_s(p)=\frac{g^2}{4\pi}D(p)[Z(p)]^2,     
\end{equation}
noting that
\begin{equation}
     Z(p) = 1, \  D(p)=\sum_{n=0}^\infty B_n\frac{p^2}{p^2-m^2_n+i\epsilon},
\end{equation}
for the dressing functions of the ghost and gluon propagators respectively, it is not difficult to verify that
\begin{equation}
     \alpha_s(p)=\frac{g^2}{4\pi}p^2f(p)
\end{equation}
with $f(0)\ne 0$. With this definition, the running coupling is seen to go to zero, at lower momenta, as $p^2$. This confirms that Yang-Mills theory is trivial in this limit. It should be emphasized that this is not true for QCD due to the presence of quarks. This kind of behavior is also seen in supersymmetric QCD where the beta function of the theory is exactly known \cite{zak}. Here, we can get the beta function for the theory from the scaling of the gluon propagator. One gets
\begin{equation}
    \mu\frac{\partial G(p)}{\partial\mu}-4\lambda\frac{\partial G(p)}{\partial\lambda}-2\gamma G(p)=0
\end{equation}
being $\gamma=0$. So, $\beta(\lambda)=4\lambda$ at the trivial fixed point. This is in agreement with recent analysis \cite{sus,pod}. So, from the mapping theorem one has $\beta(g)=4Ng^2$ and the 't Hooft coupling is seen to go to zero as $p^4$ instead, in agreement with a picture of the vacuum of the Yang-Mills theory as an instanton liquid \cite{bou2}.

\section{Comparison with lattice}

Finally, we compare the gluon propagator in the Landau gauge with lattice data. We get the following results:
\begin{figure}[H]
\begin{center}
\includegraphics[width=400pt,height=200pt]{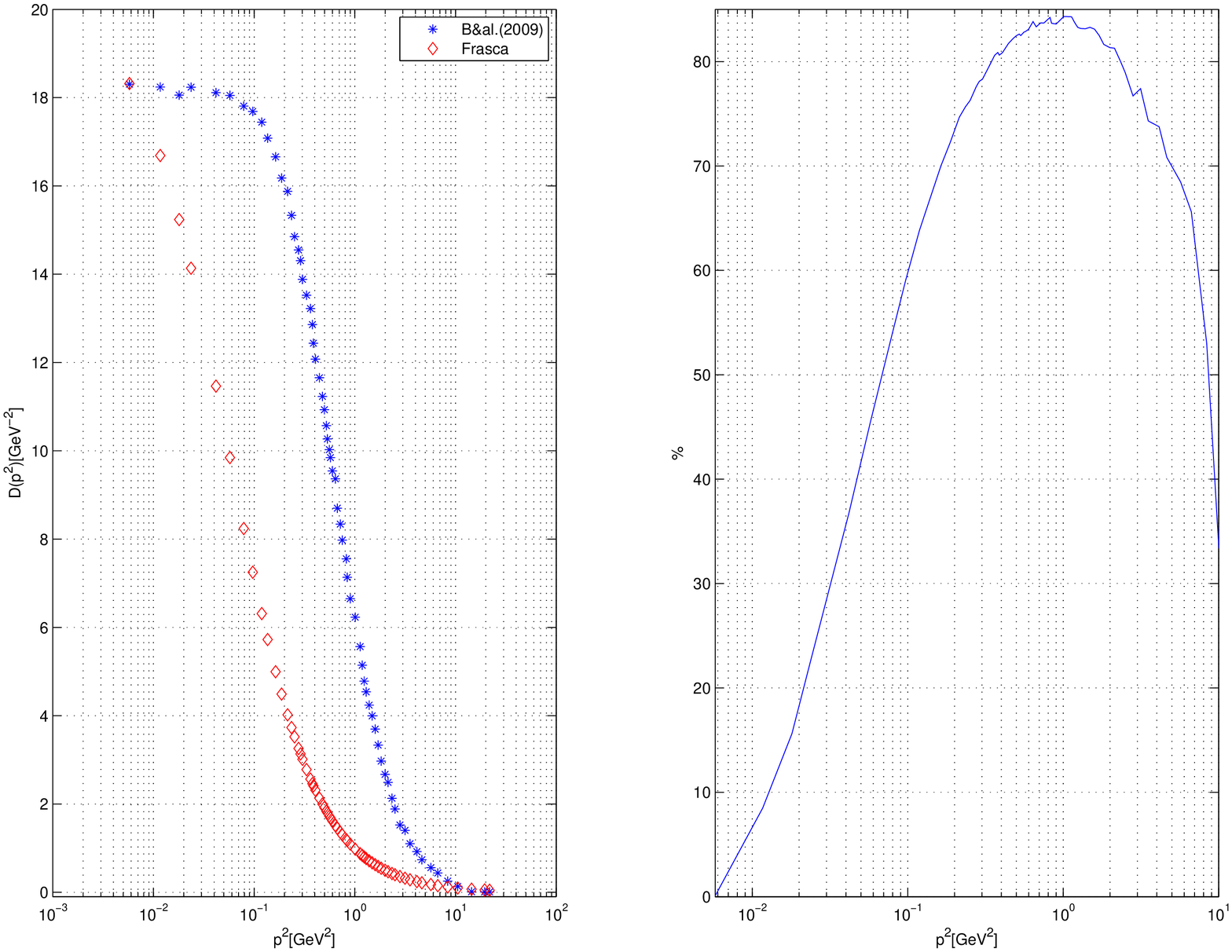}
\caption{Ref. \cite{ilg2} - SU(3) $\rm (16fm)^4$ - $\rm m_0=236\ MeV$}
\end{center}
\end{figure}

\begin{figure}[H]
\begin{center}
\includegraphics[width=400pt,height=200pt]{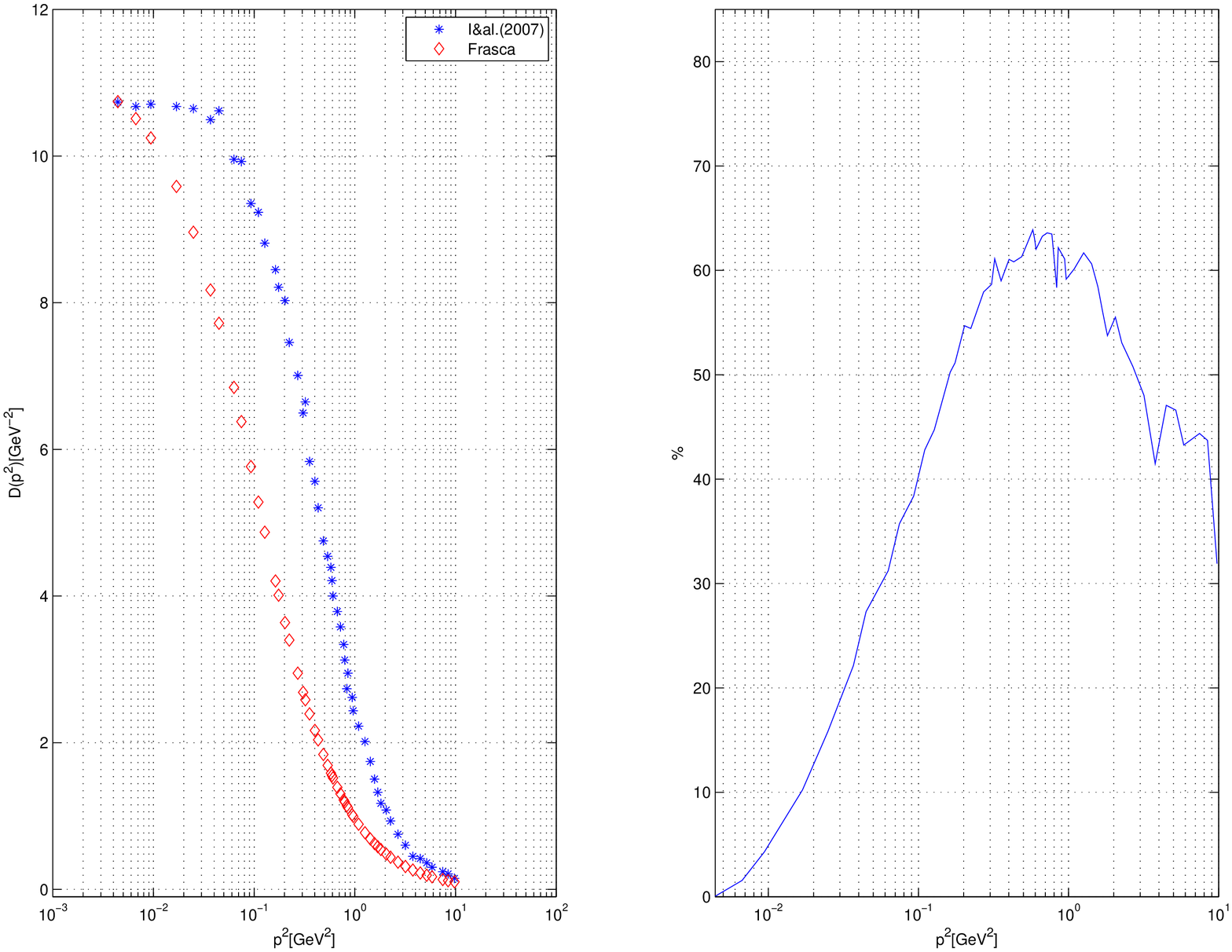}
\caption{Ref. \cite{ste2} - SU(2) $\rm (19.2fm)^4$ - $\rm m_0=317\ MeV$}
\end{center}
\end{figure}

\begin{figure}[H]
\begin{center}
\includegraphics[width=400pt,height=200pt]{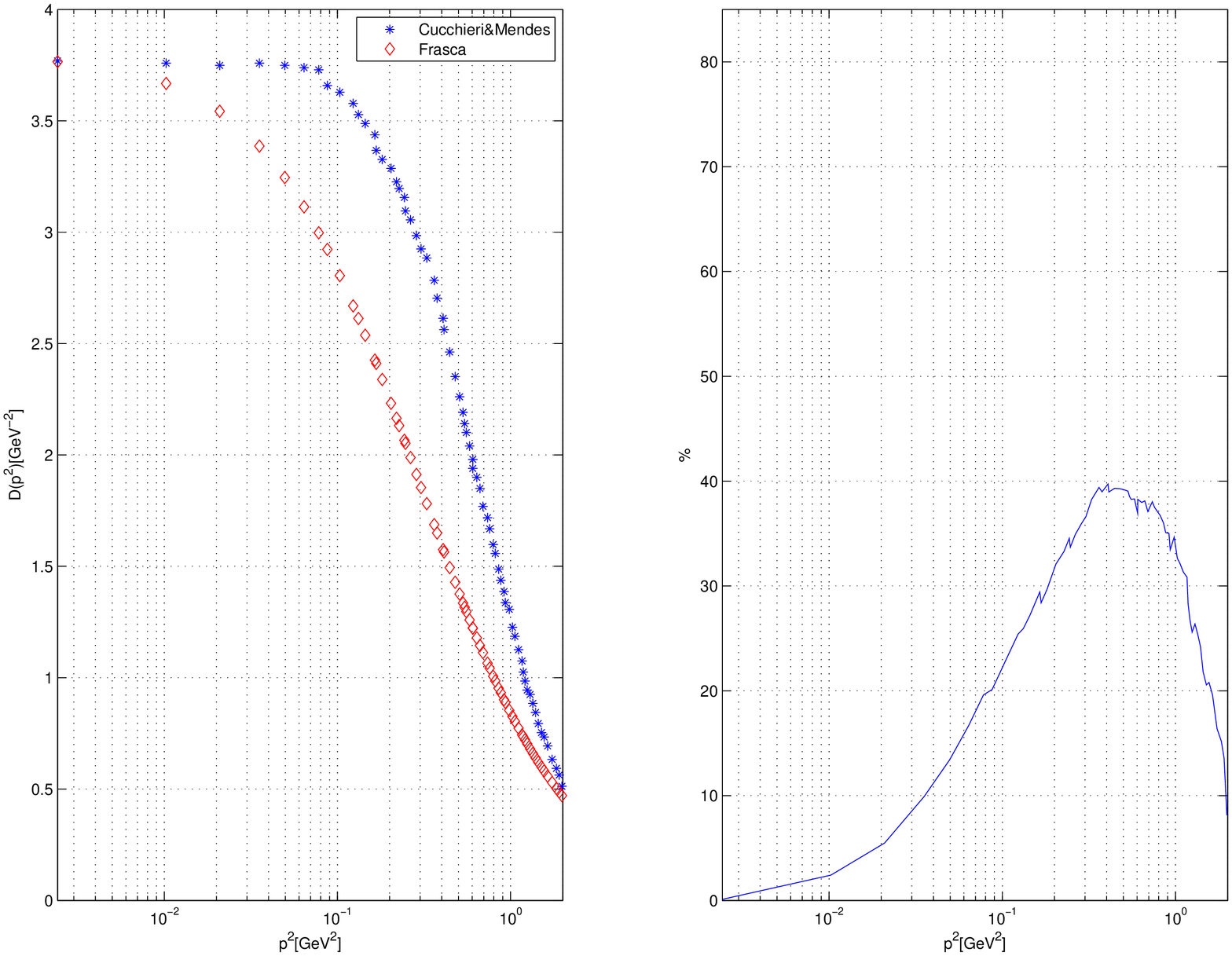}
\caption{Ref. \cite{cuc} - SU(3) $\rm (27fm)^4$ - $\rm m_0=545\ MeV$}
\end{center}
\end{figure}

From this we realize immediately that the agreement improves increasing the volume of the lattice becoming really striking with the computation of Cucchieri and Mendes \cite{cuc}. But we note also, as expected, that the error becomes larger in the intermediate region as also happens with numerical Dyson-Schwinger equations \cite{an,abp}. Finally, the dependence on the gauge group is lost into the gluon mass at this order \cite{ste2}. Gluon mass is different for each group but this is not relevant. Anyhow a value of the string tension about 400 to 440 MeV is expected and all should agree on this. It would be interesting just to fit the deep infrared region in the same way as done by Oliveira \cite{oli2} to get the full spectrum of the theory as in \cite{tep,morn} at such large volumes. Indeed, from the definition of two-point function one has
\begin{equation}
   D_{\mu\nu}^{ab}(t-t',0)=\langle TA_\mu^a(t,0)A_\nu^b(t',0)\rangle.
\end{equation}
So, using the mapping theorem one has immediately
\begin{equation}
   D_{\mu\nu}^{ab}(t-t',0)=\eta^a_\mu\eta^b_\nu\sum_{n=0}^\infty B_ne^{-im_n(t-t')}
\end{equation}
from which we can read immediately the spectrum of the theory as done in lattice computation. We have as expected $m_n=(2n+1)\frac{\pi}{2K(i)}\sqrt{\sigma}$ being $\sqrt{\sigma}=\left(Ng^2/2\right)^{\frac{1}{4}}\Lambda$ the string tension to be fixed experimentally. We can compare this spectrum with lattice computations \cite{tep} with respect to the set of pure numbers
\[
      \pi_n=\frac{m_n}{\sqrt{\sigma}}.
\]
Agreement is fairly good as seen from the following table for SU(3)
\begin{center}
\begin{table}[H]
\begin{tabular}{|c|c|c|c|} \hline\hline
Excitation & Lattice & Theoretical & Error \\ \hline
$\sigma$   & -       & 1.198140235 & - \\ \hline 
0$^{++}$   & 3.55(7) & 3.594420705 & 1\% \\ \hline
0$^{++*}$  & 5.69(10)& 5.990701175 & 5\% \\ \hline
$\sigma^*$   & -       & 2.396280470 & - \\ \hline 
2$^{++}$   & 4.78(9) & 4.792560940 & 0.2\% \\ \hline
2$^{++*}$  & - & 7.188841410 & - \\ \hline\hline
\end{tabular}
\end{table}
\end{center}
We expect the missing states to appear at higher volumes and by fitting the gluon propagator as stated above. For the other states the agreement is really striking.

\section{Conclusions}

We were able to provide a strong coupling expansion that holds both for classical and quantum field theory. The important conclusion to be drawn from this approach is that the decoupling solution is the one obtained when applied to Yang-Mills theory. Very good agreement with numerical solutions of Dyson-Schwinger equations and lattice computations give a strong support to the correctness of this technique. A set of states of free particle are obtained to obtain a quantum field theory describing the low-energy limit of Yang-Mills theory. Higher order corrections can be computed giving a propagator with a mass running with momenta. Finally, we expect to see the spectrum of the theory extracted from the propagator in the deep infrared limit. This would be a striking and unexpected merging of different lattice computations.


\begin{thebibliography}{99}
\bibitem{ilg} I. Bogolubsky, E.M. Ilgenfritz, M. Muller-Preussker and A. Sternbeck, PoS(LAT2007)290.
\bibitem{cuc} A. Cucchieri, T. Mendes, PoS(LATTICE 2007)297.
\bibitem{oli} O. Oliveira, P. J. Silva, E.-M. Ilgenfritz, A. Sternbeck, PoS(LAT2007)323.
\bibitem{alk} L. von Smekal, A. Hauck, R. Alkofer,  Annals Phys. 267 (1998) 1; Erratum-ibid. 269 (1998) 182.
\bibitem{corn} J. M. Cornwall, Phys. Rev. D 26, 1453 (1982). 
\bibitem{an} A. C. Aguilar, A. A. Natale, JHEP 0408:057 (2004). 
\bibitem{ap} A. C. Aguilar, J. Papavassiliou, JHEP 0612:012 (2006).
\bibitem{ab} D. Binosi, J. Papavassiliou, JHEP 0703:041 (2007).
\bibitem{bou} Ph. Boucaud, J.P. Leroy, A. Le Yaouanc, A.Y. Lokhov, J. Micheli, O. P\`ene, J. Rodriguez-Quintero, C. Roiesnel, arXiv:hep-ph/0507104v4.
\bibitem{fra1} M. Frasca, Phys. Rev. D 73, 027701 (2006).
\bibitem{fra2} M. Frasca, Phys. Lett. B670, 73 (2008). 
\bibitem{fra3} M. Frasca, Mod. Phys. Lett. A 24, 2425 (2009).
\bibitem{fra4} M. Frasca, Phys. Rev. A 58, 3439 (1998).
\bibitem{fra5} M. Frasca, arXiv:0802.1183v4 [hep-th].
\bibitem{smi} A. V. Smilga, {\sl Lectures in Quantum Chromodynamics}, (World Scientific, Singapore, 2001).
\bibitem{rob} R. T. Cahill and C. D. Roberts, Phys. Rev. D 32, 2419 (1985).
\bibitem{gol} T. Goldman, R. W. Haymaker, Phys. Rev. D 24, 724 (1981).
\bibitem{tao} T. Tao, private communication (2009) and \url{http://tosio.math.utoronto.ca/wiki/index.php/Talk:Yang-Mills_equations}. 
\bibitem{fra6} M. Frasca, arXiv:1007.4479v2 [hep-ph], to appear in Nuclear Physics B (Proc. Suppl.).
\bibitem{abp} A. C. Aguilar, D. Binosi, J. Papavassiliou, Phys. Rev. D 78, 025010 (2008).
\bibitem{oli2} O. Oliveira, P. J. Silva, P. Bicudo, these proceedings, arXiv:1101.5983v1 [hep-lat].
\bibitem{tep} B. Lucini, M. Teper, U. Wenger, JHEP 06:012 (2004).
\bibitem{morn} Y. Chen, A. Alexandru, S.J. Dong, T. Draper, I. Horvath, F.X. Lee, 
K.F. Liu, N. Mathur, C. Morningstar, M. Peardon, S. Tamhankar, B.L. Young, J.B. Zhang, Phys. Rev. D 73, 014516 (2006).
\bibitem{cuc2} A. Cucchieri, T. Mendes,  Phys. Rev. D 81, 016005 (2010).
\bibitem{ste} L. von Smekal, K. Maltman, A. Sternbeck, Phys. Lett. B681, 336 (2009). 
\bibitem{zak} V. A. Novikov, M. A. Shifman, A. I. Vainshtein, V. I. Zakharov, Nucl. Phys. B 229, 381 (1983). 
\bibitem{sus} I. M. Suslov, JETP 111, 450 (2010). 
\bibitem{pod} D. Podolsky, arXiv:1003.3670v1 [hep-th].
\bibitem{bou2} Ph. Boucaud, F. De Soto, A. Le Yaouanc, J.P. Leroy, J. Micheli, H. Moutarde, O. P\`ene, J. Rodríguez-Quintero, JHEP 0304:005 (2003).
\bibitem{ilg2} I.L. Bogolubsky, E.-M. Ilgenfritz, M. M\"uller-Preussker, A. Sternbeck, Phys. Lett. B676, 69 (2009).
\bibitem{ste2} A. Sternbeck, L. von Smekal, D. Leinweber and A. Williams, PoS(LAT2007)340.
\end{thebibliography}
\end{document}